\documentclass[twocolumn,showpacs,superscriptaddress,preprintnumbers,amsmath,
amssymb,prb]{revtex4}

\usepackage{graphicx}
\usepackage{dcolumn}
\usepackage{bm}
\usepackage{hyperref}

\begin{document}

\title{Evidence for magnetic clusters in BaCoO$_{3}$
}

\author{V. Pardo}
 \email{vpardo@usc.es}
\affiliation{
Departamento de F\'{\i}sica Aplicada, Facultad de F\'{\i}sica, Universidad
de Santiago de Compostela, E-15782 Campus Sur s/n, Santiago de Compostela,
Spain
}
\affiliation{
Instituto de Investigaciones Tecnol\'ogicas, Universidad de Santiago de
Compostela, E-15782, Santiago de Compostela, Spain
}

\author{J. Rivas}
\affiliation{
Departamento de F\'{\i}sica Aplicada, Facultad de F\'{\i}sica, Universidad
de Santiago de Compostela, E-15782 Campus Sur s/n, Santiago de Compostela,
Spain
}

\author{D. Baldomir}
\affiliation{
Departamento de F\'{\i}sica Aplicada, Facultad de F\'{\i}sica, Universidad
de Santiago de Compostela, E-15782 Campus Sur s/n, Santiago de Compostela,
Spain
}
\affiliation{
Instituto de Investigaciones Tecnol\'ogicas, Universidad de Santiago de
Compostela, E-15782, Santiago de Compostela, Spain
}

\author{M. Iglesias}
\affiliation{
Departamento de F\'{\i}sica Aplicada, Facultad de F\'{\i}sica, Universidad
de Santiago de Compostela, E-15782 Campus Sur s/n, Santiago de Compostela,
Spain
}
\affiliation{
Instituto de Investigaciones Tecnol\'ogicas, Universidad de Santiago de
Compostela, E-15782, Santiago de Compostela, Spain
}

\author{P. Blaha}
\affiliation{
Institute for Materials Chemistry, Vienna University of Technology, 
Getreidemarkt 9/165, A-1060 Vienna, Austria}

\author{K. Schwarz}
\affiliation{
Institute for Materials Chemistry, Vienna University of Technology, 
Getreidemarkt 9/165, A-1060 Vienna, Austria}

\author{J.E. Arias}
\affiliation{
Instituto de Investigaciones Tecnol\'ogicas, Universidad de Santiago de
Compostela, E-15782, Santiago de Compostela, Spain
}

\begin{abstract}

Magnetic properties of the transition metal oxide BaCoO$_3$ are analyzed
on the basis of the experimental and theoretical literature available via
ab inito calculations.
These can be explained by assuming the material to be formed by
noninteracting ferromagnetic clusters of about 1.2 nm in diameter separated by 
about 3 diameters. 
Above about 50 K, the so-called blocking temperature, superparamagnetic
behavior of the magnetic clusters 
occurs and, above 250 K, paramagnetism sets in. 

\end{abstract}


\maketitle

Transition metal oxides are among the most studied materials in recent
literature. The variety of their physical properties has made them
a topic of intensive research during the last few years, the phenomena of
colossal magnetoresistance \cite{cmr} and high temperature superconductivity 
\cite{sc} being the most remarkable. Among them, Co
oxides have drawn much attention, since they posses both types of
phenomena \cite{co_cmr,co_sc}. The magnetic properties of such
oxides are a matter of ongoing work. The appearance
of properties such as orbital ordering\cite{kugel}, charge
ordering\cite{chargeorder}, spin glass behavior or phase
segregation\cite{tona,mira97,spinglass,mira01,phasesegreg} makes them very rich 
in phenomena, whose understanding is a challenge for the scientific community.

BaCoO$_3$ is a transition metal oxide that recently has been a matter of 
interest of both, experimental and theoretical work
\cite{yamaura,yamaura_ssc,felser,cacheiro,vpardo_ssc,vpardo_prb}. 
Its structure\cite{acta} can be described as a 2H-hexagonal pseudoperovskite
formed by chains of distorted, tilted, and face-sharing CoO$_{6}$ octahedra.
The plane perpendicular to the chains consists of a hexagonal array of Co
atoms, which would lead to frustration for any in-plane collinear 
antiferromagnetism. The only possible long-range antiferromagnetic state 
(for collinear moments) could couple ferromagnetic
planes antiferromagnetically (so-called A-type structure)\cite{vpardo_prb}.
The magnetic properties of BaCoO$_3$ have still not been resolved in 
literature. It is certain that the Co$^{4+}$ ions are in a low-spin state 
(S=1/2, t$_{2g}^{5}$e$_{g}^{0}$), since an ion with such a high valency
 produces a large crystal field. This has been shown
both experimentally\cite{yamaura} and
theoretically\cite{felser,cacheiro,vpardo_prb}. Moreover, the
experimentally found semiconducting behavior\cite{yamaura} has been predicted
theoretically\cite{vpardo_prb}, using the LDA+U approach, to be a consequence
of an orbital ordering phenomenon, which is not yet confirmed experimentally.  
The actual magnetic configuration of the system is still uncertain. So far 
no neutron diffraction measurements are available in literature and the 
experiments that have measured the susceptibility of the system\cite{yamaura} 
are not conclusive. Theoretical
APW+lo LDA+U calculations\cite{vpardo_prb} predict a ferromagnetic state
as the magnetic ground state. The most stable
antiferromagnetic state is higher in energy but this energy difference is
extremely small and depends on the value of U. Therefore any magnetic
coupling that might occur will have a very small stabilization energy. 
This feature is typical for systems showing a cusp in their 
susceptibility\cite{review_spinglass}, as for
BaCoO$_3$.\cite{yamaura}  

In this paper, we try to shed some light in the explanation of the
magnetic properties of BaCoO$_3$ and resolve an open question in literature.
As a start we will analyze the experimental data
available and propose a model based on the existence of noninteracting 
ferromagnetic clusters in the
system, which can explain the observations. Then, using ab initio 
calculations, we will predict the size and density of the
clusters, establishing a picture which is fully consistent with both
experimental and theoretical data.

When we take a close look at Ref. \onlinecite{yamaura}, Fig. 10, we observe
susceptibility curves that are typical for a ``fine-particle" 
system\cite{fiorani} with a blocking temperature of around 50 K. 
This is approximately, where the field cooling (FC) magnetization curve 
separates from the zero-field cooling (ZFC) one for an applied field of 1 kOe. 
These FC and ZFC curves indicate that BaCoO$_{3}$ is formed by regions or 
clusters, whose magnetic moments are ferromagnetically ordered, but are 
dispersed into a non-ferromagnetic matrix, a situation that can be 
described by Wohlfarth's superparamagnetic model\cite{wohlfarth}.
The shape of the FC curve below the
blocking temperature indicates that the clusters are widely separated from
each other and they do not interact strongly. However, would they interact, 
this curve would be flat instead of growing exponentially as is 
observed\cite{fiorani}.

The blocking temperature depends on the size of the clusters, the
anisotropy constant of the material, the applied field and on the measuring 
time (on the apparatus utilized for the experiment). 
By analyzing the data from Ref. \onlinecite{yamaura}, one can see that the
blocking temperature decreases when the applied field increases. This is
the typical behavior of fine-particle systems, even though the contrary
might occur if the clusters grow with the applied field \cite{rivadulla}.
Based on the theory developed by N\'eel
for superparamagnetic particles\cite{neel,wolfarth}, we can predict that the
highest blocking temperature (which is reached when the applied field is
much smaller than the so-called anisotropy field) is basically obtained at
1kOe. According to this, the mean blocking temperature varies as:

\begin{equation}
\left(\frac{T_B(H)}{T_B(0)}\right)^{1/2}=1-\frac{H}{H_K}
\end{equation}

where $T_B$ is the blocking temperature and $H_K$ is the mean anisotropy field.
From the data in Ref. \onlinecite{yamaura}, the blocking temperature is
approximately 10 K for H=10kOe and 50 K for H=1kOe. From this, the maximum
blocking temperature that can be reached at a very small field is about 53
K, a limit that is basically obtained at a field of 1kOe. This is caused by the
large anisotropy of the system, which is due to the quasi-one-dimensional 
structure formed by widely separated chains of Co atoms. Hence, 
the value we have chosen to explain the magnetic properties of the
system is T$_B$$\simeq$ 50 K, the expected value for low 
fields. The description of the system, which will be given below, 
corresponds to these low applied fields.

From the value of the blocking temperature we can estimate the mean size of the
clusters. For doing so, we have calculated the anisotropy constant of the 
system from ab initio methods.
The calculations were performed using the WIEN2k\cite{wien,wien2k}
software, a package that uses a full-potential, all electron
APW+lo\cite{sjo} method
that allows to carry out total energy calculations with one of the most
precise methods available. For
this moderately correlated transition metal oxide, we used the LDA+U
approach including self-interaction corrections\cite{sic2} in the so-called 
``fully-localized limit"\cite{mazin}
with U= 5eV and J= 0.5eV, values discussed in Ref. \onlinecite{vpardo_prb}. This
method has proven reliable for transition metal oxides\cite{anisimov},
since it improves over GGA (generalized gradient approximation) or LDA
(local density approximation) in the study of correlated electrons by
means of introducing the on-site Coulomb repulsion U. The non-orbital
dependent part of the exchange-correlation potential was calculated using
the GGA in the PBE (Perdew-Burke-Ernzerhof) scheme\cite{gga}.
Local orbitals (Co 3s, 3p, 3d, Ba 4d, 5s, 5p, O 2s and 2p) were added to
obtain a better flexibility of the basis set and to improve the description of 
the semicore states. R$_{mt}$K$_{max}$=7 and 500 k-points in the
irreducible Brillouin zone (1200 k-points in the whole Brillouin zone)
were taken. For calculating the magnetic anisotropy, a fully relativistic
calculation was carried out, where spin-orbit effects
were included using a second variational scheme\cite{singh}.

For calculating the energy contribution of the magnetic anisotropy we
studied the system in different configurations with the magnetic moments 
along the different crystallographic directions: (100), (110) 
and (111), and compared their total energies.
The system was studied in a ferromagnetic low-spin state with
an alternating orbital order along the c axis as described in
Ref. \onlinecite{vpardo_prb}, which in this paper was shown to be the 
ground state.
From our total energy calculations, the value of the anisotropy constant
is 1.2 mRy per unit cell, which contains 2 Co atoms (about
2$\times$10$^8$erg/cm$^3$). This value is
larger than the anisotropy energy of a system of fine Co particles
($\simeq$ 7$\times$10$^6$erg/cm$^3$). We are not
considering the shape anisotropy due to the formation of
clusters, but 
we can assume this contribution to be a secondary effect in this highly
anisotropic compound.
The lowest energy state has the moments lying in the hexagonal plane
(along the b axis). We assume the system to have cubic anisotropy (since 
the Co ions are in an octahedral environment) and found the value of the
anisotropy constant as $K=\frac{E_B}{4}$, where $E_B$ is the energy
difference between the easy (b) and hard (c) axis.
The procedure for estimating the volume of each cluster is via the following
formula for superparamagnetic particles\cite{morrish}: 
\begin{equation}\label{kvequ}
KV= 25k_BT_B
\end{equation}
where
$T_B$ is the blocking temperature, $V$ is the mean volume of the cluster and
$k_B$ is the Boltzmann constant. The 
factor 25 comes from the measuring
time, which was assumed with a time constant $\tau$= 10$^2$ s for the experiments 
of magnetic susceptibility under FC and ZFC
conditions in Ref. \onlinecite{yamaura}. 
Using equation \ref{kvequ} with the anisotropy
constant calculated ab initio and the blocking temperature from the
experimental data in Ref. \onlinecite{yamaura} (T$_B\simeq$ 50 K), we estimate
that each cluster has a typical diameter of approximately 1.2 nm, containing 
about 14 Co atoms. The value of the anisotropy constant was calculated at T= 0,
hence the volume of the cluster must be considered as a lowest limit. There 
exists a distribution of particle sizes. This is evident
by the fact that the maximum of the ZFC curve is displaced from the point
where ZFC and FC curves separate from one another. In our description, we
will stick to the average values.

From the blocking temperature (at about 50 K) to approximately 250
K, the system shows a superparamagnetic behavior, entering the normal
paramagnetic regime at about 250 K. This superparamagnetism can be due to
the presence of magnetic clusters, which can be identified as an assembly
of single domain particles with a total magnetic moment of about 14
$\mu_B$ per particle
(from the ab initio calculations in Ref. \onlinecite{vpardo_prb},
the total magnetic moment per Co atom is 1.00 $\mu_B$), where the magnetization 
of the sample follows the Langevin formula: 

\begin{equation}
M = N\mu L(y)  
\end{equation}

where $N$ is the density of clusters per unit volume, $\mu$ is the magnetic moment 
of each
cluster, $y=\frac{\mu H}{k_BT}$ and $L(y)$ is the Langevin function. At
small $y$, the magnetization varies linearly with $H/T$ and the Langevin
function approximates to a Curie law:
\begin{equation}
M \simeq N \frac{\mu^2H}{3k_BT}
\end{equation}

\begin{figure}
\includegraphics{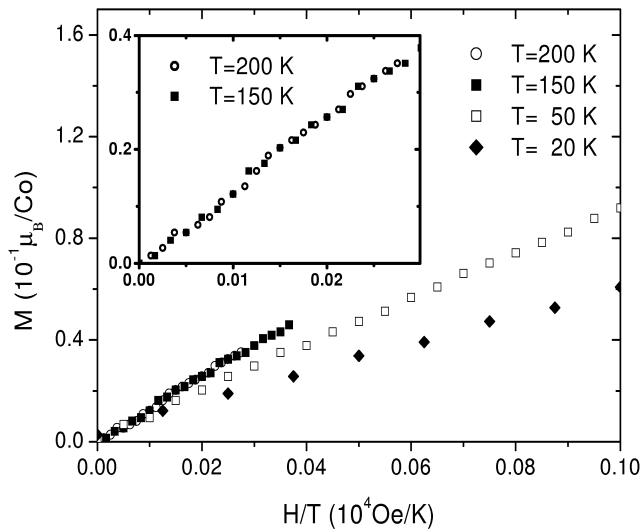}
\caption{Total magnetization vs. H/T for four values of the temperature (data
from Yamaura \textsl{et al.}\cite{yamaura}). In the inset, the curves at
T= 150 K and T= 200 K are highlighted, showing an excellent H/T
superposition.
However, for T$\leq$ 50 K, straight lines have a smaller slope due to the
anisotropy effects at temperatures comparable to the blocking temperature.
}\label{mvsht}
\end{figure}

Figure \ref{mvsht} shows the data from Ref. \onlinecite{yamaura} as $M$
vs. $H/T$. This gives a straight line, that corresponds to the Langevin
formula (superparamagnetism) at ``normal" temperatures (50$\leq$T(K)$\leq$200)
and ``moderate" fields (0$\leq$H(Oe)$\leq$5$\times$10$^4$). But, 
the same type of plot has a different slope for T= 50 K and T= 20 K,
because they are below the blocking temperature 
and the clusters are said to be ``blocked" (the effects due to the
blocking are appreciable in a region around the blocking temperature,
superparamagnetism is observed where ZFC and FC curves match each other,
above some 70 K, from the curves in Ref. \onlinecite{yamaura}). 
The slope increases with T and approaches a maximum when T $>$$>$ T$_B$, being
the difference remarkable in the vicinity of T$_B$, where thermal
equilibrium disappears. 

From these curves we have estimated the density of clusters that exists in
BaCoO$_3$, being the result that we have one 1.2 nm-diameter cluster in a
spherical volume with a diameter
of about 3.0 nm. Hence, the average distance between two clusters is about 3
diameters, big enough to assume that they do not strongly interact with
each other. As we
mentioned above, this fact coincides with the shape of the curves in Ref.
\onlinecite{yamaura} that ressemble noninteracting superparamagnetic
particles\cite{fiorani}. 

The normal paramagnetic state appears above some 250 K, when thermal
energy is enough to break the magnetic order of the crystal. 
Below the blocking temperature, T$_B\simeq$ 50 K, magnetic moments within the 
cluster lie along the easy axis and for this system, formed by
ferromagnetic clusters, hysteresis loops are expected.

One of the possible mechanism for the formation of these clusters is Nagaev's
theory\cite{nagaev,ivan}, according to which a conduction electron might become
trapped and polarizes ferromagnetically the antiferromagnetic medium giving 
rise to the
occurrence of ferromagnetic particles dispersed into a non-ferromagnetic
matrix, producing a phase separation phenomenon.

In this paper, we have tried to explain the magnetic properties of the transition
metal oxide BaCoO$_3$ utilizing ab initio full potential APW+lo LDA+U 
calculations to
interpret the experimental data available. The theoretical prediction of
ferromagnetism as the ground state does not coincide with the experimental
data of magnetic susceptibility, which is not that of a long-range ferromagnet. 
The explanation we propose is based on the formation of noninteracting 
ferromagnetic clusters 
in the system. Using ab initio calculations, we have estimated the
distribution of these clusters. They have a typical size of about 1.2 nm in 
diameter, and are
separated by about 3 diameters, which explains the evidences that they do
not interact strongly. With this model, the susceptibility curves can be
explained, corresponding to a superparamagnetic behavior in the
temperature region from the blocking temperature ($T_B\simeq$ 50 K) up to 250 K. 
Also, the change in the slope of the magnetization vs. H/T curves below 50
K is related to the blocking of the superparamagnetic state.
Above 250 K, normal paramagnetism is found.

From our conclusions, BaCoO$_3$ is expected to be an example of a system
with a similar behavior to that formed by
``ultra-fine" noninteracting magnetic particles and further
experimental work is needed to provide more insight into the
characterization of the material. SANS (small angle neutron scattering)
measurements and in addition a better static and dynamic study of the variation
of the magnetization with temperature and applied field is strongly encouraged.

We wish to thank F. Rivadulla, M.N. Baibich and K.I. Kugel for a fruitful 
discussion of the manuscript. The authors also wish to thank the CESGA (Centro 
de Computaci\'on de Galicia) for the computing facilities, the Xunta de Galicia 
for the financial support through a PhD grant and through the project
PGIDIT02TMT20601PR and the support from the DGI of the Ministry of Science 
and Technology of Spain under the project FEDER MAT 2001-3749.



\begin{thebibliography}{35}
\expandafter\ifx\csname natexlab\endcsname\relax\def\natexlab#1{#1}\fi
\expandafter\ifx\csname bibnamefont\endcsname\relax
  \def\bibnamefont#1{#1}\fi
\expandafter\ifx\csname bibfnamefont\endcsname\relax
  \def\bibfnamefont#1{#1}\fi
\expandafter\ifx\csname citenamefont\endcsname\relax
  \def\citenamefont#1{#1}\fi
\expandafter\ifx\csname url\endcsname\relax
  \def\url#1{\texttt{#1}}\fi
\expandafter\ifx\csname urlprefix\endcsname\relax\def\urlprefix{URL }\fi
\providecommand{\bibinfo}[2]{#2}
\providecommand{\eprint}[2][]{\url{#2}}

\bibitem[{\citenamefont{Kusters et~al.}(1989)\citenamefont{Kusters, Singleton,
  Keen, McGreevy, and Hages}}]{cmr}
\bibinfo{author}{\bibfnamefont{R.~M.} \bibnamefont{Kusters}},
  \bibinfo{author}{\bibfnamefont{J.}~\bibnamefont{Singleton}},
  \bibinfo{author}{\bibfnamefont{D.~A.} \bibnamefont{Keen}},
  \bibinfo{author}{\bibfnamefont{R.}~\bibnamefont{McGreevy}}, \bibnamefont{and}
  \bibinfo{author}{\bibfnamefont{W.}~\bibnamefont{Hages}},
  \bibinfo{journal}{Appl. Phys. Lett.} \textbf{\bibinfo{volume}{63}},
  \bibinfo{pages}{1990} (\bibinfo{year}{1989}).

\bibitem[{\citenamefont{Bednorz and Muller}(1986)}]{sc}
\bibinfo{author}{\bibfnamefont{J.~G.} \bibnamefont{Bednorz}} \bibnamefont{and}
  \bibinfo{author}{\bibfnamefont{K.~A.} \bibnamefont{Muller}},
  \bibinfo{journal}{Z. Phys. B} \textbf{\bibinfo{volume}{64}},
  \bibinfo{pages}{189} (\bibinfo{year}{1986}).

\bibitem[{\citenamefont{Martin et~al.}(1997)\citenamefont{Martin, Maignan,
  Pelloquin, Nguygen, and Raveau}}]{co_cmr}
\bibinfo{author}{\bibfnamefont{C.}~\bibnamefont{Martin}},
  \bibinfo{author}{\bibfnamefont{A.}~\bibnamefont{Maignan}},
  \bibinfo{author}{\bibfnamefont{D.}~\bibnamefont{Pelloquin}},
  \bibinfo{author}{\bibfnamefont{N.}~\bibnamefont{Nguygen}}, \bibnamefont{and}
  \bibinfo{author}{\bibfnamefont{B.}~\bibnamefont{Raveau}},
  \bibinfo{journal}{Appl. Phys. Lett.} \textbf{\bibinfo{volume}{71}},
  \bibinfo{pages}{1421} (\bibinfo{year}{1997}).

\bibitem[{\citenamefont{Takada et~al.}(2003)\citenamefont{Takada, Sakurai,
  Takayama-Muromachi, Izumi, Dilanian, and Sasaki}}]{co_sc}
\bibinfo{author}{\bibfnamefont{K.}~\bibnamefont{Takada}},
  \bibinfo{author}{\bibfnamefont{H.}~\bibnamefont{Sakurai}},
  \bibinfo{author}{\bibfnamefont{E.}~\bibnamefont{Takayama-Muromachi}},
  \bibinfo{author}{\bibfnamefont{F.}~\bibnamefont{Izumi}},
  \bibinfo{author}{\bibfnamefont{R.}~\bibnamefont{Dilanian}}, \bibnamefont{and}
  \bibinfo{author}{\bibfnamefont{T.}~\bibnamefont{Sasaki}},
  \bibinfo{journal}{Nature (London)} \textbf{\bibinfo{volume}{422}},
  \bibinfo{pages}{53} (\bibinfo{year}{2003}).

\bibitem[{\citenamefont{Kugel and Khomskii}(1973)}]{kugel}
\bibinfo{author}{\bibfnamefont{K.}~\bibnamefont{Kugel}} \bibnamefont{and}
  \bibinfo{author}{\bibfnamefont{D.}~\bibnamefont{Khomskii}},
  \bibinfo{journal}{Sov. Phys.-JETP} \textbf{\bibinfo{volume}{37}},
  \bibinfo{pages}{725} (\bibinfo{year}{1973}).

\bibitem[{\citenamefont{Cheong and Hwang}(1999)}]{chargeorder}
\bibinfo{author}{\bibfnamefont{S.}~\bibnamefont{Cheong}} \bibnamefont{and}
  \bibinfo{author}{\bibfnamefont{H.}~\bibnamefont{Hwang}},
  \emph{\bibinfo{title}{Contribution to Colossal Magnetoresistance Oxides,
  Monographs in Condensed Matter Science, Edited by Y. Tokura}}
  (\bibinfo{address}{Gordon, Breach, London}, \bibinfo{year}{1999}).

\bibitem[{\citenamefont{Mira et~al.}(1999)\citenamefont{Mira, Rivas, Jonason,
  Nordblad, Breijo, and Se\~nar\'{\i}s Rodr\'{\i}guez}}]{spinglass}
\bibinfo{author}{\bibfnamefont{J.}~\bibnamefont{Mira}},
  \bibinfo{author}{\bibfnamefont{J.}~\bibnamefont{Rivas}},
  \bibinfo{author}{\bibfnamefont{K.}~\bibnamefont{Jonason}},
  \bibinfo{author}{\bibfnamefont{P.}~\bibnamefont{Nordblad}},
  \bibinfo{author}{\bibfnamefont{M.}~\bibnamefont{Breijo}}, \bibnamefont{and}
  \bibinfo{author}{\bibfnamefont{M.}~\bibnamefont{Se\~nar\'{\i}s
  Rodr\'{\i}guez}}, \bibinfo{journal}{J. Magn. Magn. Mater.}
  \textbf{\bibinfo{volume}{196-197}}, \bibinfo{pages}{487}
  (\bibinfo{year}{1999}).

\bibitem[{\citenamefont{Hoch et~al.}(2004)\citenamefont{Hoch, Kuhns, Moulton,
  Reyes, Wu, and Leighton}}]{phasesegreg}
\bibinfo{author}{\bibfnamefont{M.}~\bibnamefont{Hoch}},
  \bibinfo{author}{\bibfnamefont{P.}~\bibnamefont{Kuhns}},
  \bibinfo{author}{\bibfnamefont{W.}~\bibnamefont{Moulton}},
  \bibinfo{author}{\bibfnamefont{A.}~\bibnamefont{Reyes}},
  \bibinfo{author}{\bibfnamefont{J.}~\bibnamefont{Wu}}, \bibnamefont{and}
  \bibinfo{author}{\bibfnamefont{C.}~\bibnamefont{Leighton}},
  \bibinfo{journal}{Phys. Rev. B} \textbf{\bibinfo{volume}{69}},
  \bibinfo{pages}{014425} (\bibinfo{year}{2004}).

\bibitem[{\citenamefont{Se\~nar\'{\i}s Rodr\'{\i}guez and Goodenough}(1995)}]{tona}
\bibinfo{author}{\bibfnamefont{M.}~\bibnamefont{Se\~nar\'{\i}s Rodr\'{\i}guez}}
  \bibnamefont{and}
  \bibinfo{author}{\bibfnamefont{J.}~\bibnamefont{Goodenough}},
  \bibinfo{journal}{J. Solid State Chem.} \textbf{\bibinfo{volume}{118}},
  \bibinfo{pages}{323} (\bibinfo{year}{1995}).

\bibitem[{\citenamefont{Mira et~al.}(1997)\citenamefont{Mira, Rivas, S\'anchez,
  Se\~nar\'{\i}s Rodr\'{\i}guez, Fiorani, Rinaldi, and Caciuffo}}]{mira97}
\bibinfo{author}{\bibfnamefont{J.}~\bibnamefont{Mira}},
  \bibinfo{author}{\bibfnamefont{J.}~\bibnamefont{Rivas}},
  \bibinfo{author}{\bibfnamefont{R.}~\bibnamefont{S\'anchez}},
  \bibinfo{author}{\bibfnamefont{M.}~\bibnamefont{Se\~nar\'{\i}s
  Rodr\'{\i}guez}}, \bibinfo{author}{\bibfnamefont{D.}~\bibnamefont{Fiorani}},
  \bibinfo{author}{\bibfnamefont{D.}~\bibnamefont{Rinaldi}}, \bibnamefont{and}
  \bibinfo{author}{\bibfnamefont{R.}~\bibnamefont{Caciuffo}},
  \bibinfo{journal}{J. Appl. Phys.} \textbf{\bibinfo{volume}{81}},
  \bibinfo{pages}{5753} (\bibinfo{year}{1997}).

\bibitem[{\citenamefont{Mira et~al.}(2001)\citenamefont{Mira, Rivas, Baio,
  Caciuffo, Rinaldi, Fiorani, and Se\~nar\'{\i}s Rodr\'{\i}guez}}]{mira01}
\bibinfo{author}{\bibfnamefont{J.}~\bibnamefont{Mira}},
  \bibinfo{author}{\bibfnamefont{J.}~\bibnamefont{Rivas}},
  \bibinfo{author}{\bibfnamefont{G.}~\bibnamefont{Baio}},
  \bibinfo{author}{\bibfnamefont{R.}~\bibnamefont{Caciuffo}},
  \bibinfo{author}{\bibfnamefont{D.}~\bibnamefont{Rinaldi}},
  \bibinfo{author}{\bibfnamefont{D.}~\bibnamefont{Fiorani}}, \bibnamefont{and}
  \bibinfo{author}{\bibfnamefont{M.}~\bibnamefont{Se\~nar\'{\i}s
  Rodr\'{\i}guez}}, \bibinfo{journal}{J. Appl. Phys.}
  \textbf{\bibinfo{volume}{89}}, \bibinfo{pages}{5606} (\bibinfo{year}{2001}).

\bibitem[{\citenamefont{Yamaura et~al.}(1999)\citenamefont{Yamaura, Zandbergen,
  Abe, and Cava}}]{yamaura}
\bibinfo{author}{\bibfnamefont{K.}~\bibnamefont{Yamaura}},
  \bibinfo{author}{\bibfnamefont{H.}~\bibnamefont{Zandbergen}},
  \bibinfo{author}{\bibfnamefont{K.}~\bibnamefont{Abe}}, \bibnamefont{and}
  \bibinfo{author}{\bibfnamefont{R.}~\bibnamefont{Cava}}, \bibinfo{journal}{J.
  Solid State Chem.} \textbf{\bibinfo{volume}{146}}, \bibinfo{pages}{96}
  (\bibinfo{year}{1999}).

\bibitem[{\citenamefont{Yamaura and Cava}(2000)}]{yamaura_ssc}
\bibinfo{author}{\bibfnamefont{K.}~\bibnamefont{Yamaura}} \bibnamefont{and}
  \bibinfo{author}{\bibfnamefont{R.~J.} \bibnamefont{Cava}},
  \bibinfo{journal}{Solid State Commun.} \textbf{\bibinfo{volume}{115}},
  \bibinfo{pages}{301} (\bibinfo{year}{2000}).

\bibitem[{\citenamefont{Felser et~al.}(1999)\citenamefont{Felser, Yamaura, and
  Cava}}]{felser}
\bibinfo{author}{\bibfnamefont{C.}~\bibnamefont{Felser}},
  \bibinfo{author}{\bibfnamefont{K.}~\bibnamefont{Yamaura}}, \bibnamefont{and}
  \bibinfo{author}{\bibfnamefont{R.}~\bibnamefont{Cava}}, \bibinfo{journal}{J.
  Solid State Chem.} \textbf{\bibinfo{volume}{146}}, \bibinfo{pages}{411}
  (\bibinfo{year}{1999}).

\bibitem[{\citenamefont{Cacheiro et~al.}(2003)\citenamefont{Cacheiro, Iglesias,
  Pardo, Baldomir, and Arias}}]{cacheiro}
\bibinfo{author}{\bibfnamefont{J.~L.} \bibnamefont{Cacheiro}},
  \bibinfo{author}{\bibfnamefont{M.}~\bibnamefont{Iglesias}},
  \bibinfo{author}{\bibfnamefont{V.}~\bibnamefont{Pardo}},
  \bibinfo{author}{\bibfnamefont{D.}~\bibnamefont{Baldomir}}, \bibnamefont{and}
  \bibinfo{author}{\bibfnamefont{J.}~\bibnamefont{Arias}},
  \bibinfo{journal}{Int. J. Quantum Chem.} \textbf{\bibinfo{volume}{91}},
  \bibinfo{pages}{252} (\bibinfo{year}{2003}).

\bibitem[{\citenamefont{Pardo et~al.}(2003{\natexlab{a}})\citenamefont{Pardo,
  Iglesias, Baldomir, Castro, and Arias}}]{vpardo_ssc}
\bibinfo{author}{\bibfnamefont{V.}~\bibnamefont{Pardo}},
  \bibinfo{author}{\bibfnamefont{M.}~\bibnamefont{Iglesias}},
  \bibinfo{author}{\bibfnamefont{D.}~\bibnamefont{Baldomir}},
  \bibinfo{author}{\bibfnamefont{J.}~\bibnamefont{Castro}}, \bibnamefont{and}
  \bibinfo{author}{\bibfnamefont{J.}~\bibnamefont{Arias}},
  \bibinfo{journal}{Solid State Commun.} \textbf{\bibinfo{volume}{128}},
  \bibinfo{pages}{101} (\bibinfo{year}{2003}{\natexlab{a}}).

\bibitem[{\citenamefont{Pardo et~al.}(2003{\natexlab{b}})\citenamefont{Pardo,
  Blaha, Iglesias, Schwarz, Baldomir, and Arias}}]{vpardo_prb}
\bibinfo{author}{\bibfnamefont{V.}~\bibnamefont{Pardo}},
  \bibinfo{author}{\bibfnamefont{P.}~\bibnamefont{Blaha}},
  \bibinfo{author}{\bibfnamefont{M.}~\bibnamefont{Iglesias}},
  \bibinfo{author}{\bibfnamefont{K.}~\bibnamefont{Schwarz}},
  \bibinfo{author}{\bibfnamefont{D.}~\bibnamefont{Baldomir}}, \bibnamefont{and}
  \bibinfo{author}{\bibfnamefont{J.}~\bibnamefont{Arias}},
  \bibinfo{journal}{cond. mat.} \textbf{\bibinfo{volume}{0405082}}
  (\bibinfo{year}{2003}{\natexlab{b}}).

\bibitem[{\citenamefont{Taguchi et~al.}(1977)\citenamefont{Taguchi, Takeda,
  Kanamaru, Shimada, and Koizumi}}]{acta}
\bibinfo{author}{\bibfnamefont{H.}~\bibnamefont{Taguchi}},
  \bibinfo{author}{\bibfnamefont{Y.}~\bibnamefont{Takeda}},
  \bibinfo{author}{\bibfnamefont{F.}~\bibnamefont{Kanamaru}},
  \bibinfo{author}{\bibfnamefont{M.}~\bibnamefont{Shimada}}, \bibnamefont{and}
  \bibinfo{author}{\bibfnamefont{M.}~\bibnamefont{Koizumi}},
  \bibinfo{journal}{Acta Crystallogr. B} \textbf{\bibinfo{volume}{33}},
  \bibinfo{pages}{1299} (\bibinfo{year}{1977}).

\bibitem[{\citenamefont{Binder and Young}(1986)}]{review_spinglass}
\bibinfo{author}{\bibfnamefont{K.}~\bibnamefont{Binder}} \bibnamefont{and}
  \bibinfo{author}{\bibfnamefont{A.}~\bibnamefont{Young}},
  \bibinfo{journal}{Rev. Mod. Phys.} \textbf{\bibinfo{volume}{58}},
  \bibinfo{pages}{801} (\bibinfo{year}{1986}).

\bibitem[{\citenamefont{Dormann et~al.}(1997)\citenamefont{Dormann, Fiorani,
  and Tronc}}]{fiorani}
\bibinfo{author}{\bibfnamefont{J.}~\bibnamefont{Dormann}},
  \bibinfo{author}{\bibfnamefont{D.}~\bibnamefont{Fiorani}}, \bibnamefont{and}
  \bibinfo{author}{\bibfnamefont{E.}~\bibnamefont{Tronc}},
  \bibinfo{journal}{Adv. Chem. Phys.} \textbf{\bibinfo{volume}{98}},
  \bibinfo{pages}{283} (\bibinfo{year}{1997}).

\bibitem[{\citenamefont{Wohlfarth}(1979)}]{wohlfarth}
\bibinfo{author}{\bibfnamefont{E.}~\bibnamefont{Wohlfarth}},
  \bibinfo{journal}{Phys. Lett. A} \textbf{\bibinfo{volume}{70}},
  \bibinfo{pages}{489} (\bibinfo{year}{1979}).

\bibitem[{\citenamefont{Rivadulla et~al.}(2004)\citenamefont{Rivadulla,
  Quintela, and Rivas}}]{rivadulla}
\bibinfo{author}{\bibfnamefont{F.}~\bibnamefont{Rivadulla}},
  \bibinfo{author}{\bibfnamefont{M.~L.} \bibnamefont{Quintela}},
  \bibnamefont{and} \bibinfo{author}{\bibfnamefont{J.}~\bibnamefont{Rivas}},
  \bibinfo{journal}{cond.-mat./0403686}  (\bibinfo{year}{2004}).

\bibitem[{\citenamefont{N\'eel}(1949)}]{neel}
\bibinfo{author}{\bibfnamefont{L.}~\bibnamefont{N\'eel}},
  \bibinfo{journal}{Ann. G\'eophys.} \textbf{\bibinfo{volume}{5}},
  \bibinfo{pages}{99} (\bibinfo{year}{1949}).

\bibitem[{\citenamefont{Wohlfarth}(1980)}]{wolfarth}
\bibinfo{author}{\bibfnamefont{E.}~\bibnamefont{Wohlfarth}},
  \bibinfo{journal}{J. Phys. F: Metal Phys.} \textbf{\bibinfo{volume}{10}}
  (\bibinfo{year}{1980}).

\bibitem[{\citenamefont{Schwarz and Blaha}(2003)}]{wien}
\bibinfo{author}{\bibfnamefont{K.}~\bibnamefont{Schwarz}} \bibnamefont{and}
  \bibinfo{author}{\bibfnamefont{P.}~\bibnamefont{Blaha}},
  \bibinfo{journal}{Comp. Mat. Sci.} \textbf{\bibinfo{volume}{28}},
  \bibinfo{pages}{259} (\bibinfo{year}{2003}).

\bibitem[{\citenamefont{Blaha et~al.}(2001)\citenamefont{Blaha, Schwarz,
  Madsen, Kvasnicka, and Luitz}}]{wien2k}
\bibinfo{author}{\bibfnamefont{P.}~\bibnamefont{Blaha}},
  \bibinfo{author}{\bibfnamefont{K.}~\bibnamefont{Schwarz}},
  \bibinfo{author}{\bibfnamefont{G.~K.~H.} \bibnamefont{Madsen}},
  \bibinfo{author}{\bibfnamefont{D.}~\bibnamefont{Kvasnicka}},
  \bibnamefont{and} \bibinfo{author}{\bibfnamefont{J.}~\bibnamefont{Luitz}},
  \emph{\bibinfo{title}{WIEN2k, An Augmented Plane Wave Plus Local Orbitals
  Program for Calculating Crystal Properties. ISBN 3-9501031-1-2}},
  \bibinfo{address}{Vienna University of Technology, Austria}
  (\bibinfo{year}{2001}).

\bibitem[{\citenamefont{Sj{\"o}stedt et~al.}(2000)\citenamefont{Sj{\"o}stedt,
  N{\"o}rdstrom, and Singh}}]{sjo}
\bibinfo{author}{\bibfnamefont{E.}~\bibnamefont{Sj{\"o}stedt}},
  \bibinfo{author}{\bibfnamefont{L.}~\bibnamefont{N{\"o}rdstrom}},
  \bibnamefont{and} \bibinfo{author}{\bibfnamefont{D.}~\bibnamefont{Singh}},
  \bibinfo{journal}{Solid State Commun.} \textbf{\bibinfo{volume}{114}},
  \bibinfo{pages}{15} (\bibinfo{year}{2000}).

\bibitem[{\citenamefont{Lichtenstein et~al.}(1995)\citenamefont{Lichtenstein,
  Anisimov, and Zaanen}}]{sic2}
\bibinfo{author}{\bibfnamefont{A.}~\bibnamefont{Lichtenstein}},
  \bibinfo{author}{\bibfnamefont{V.}~\bibnamefont{Anisimov}}, \bibnamefont{and}
  \bibinfo{author}{\bibfnamefont{J.}~\bibnamefont{Zaanen}},
  \bibinfo{journal}{Phys.\ Rev. B} \textbf{\bibinfo{volume}{52}},
  \bibinfo{pages}{R5467} (\bibinfo{year}{1995}).

\bibitem[{\citenamefont{Petukhov et~al.}(2003)\citenamefont{Petukhov, Mazin,
  Chioncel, and Lichtenstein}}]{mazin}
\bibinfo{author}{\bibfnamefont{A.}~\bibnamefont{Petukhov}},
  \bibinfo{author}{\bibfnamefont{I.}~\bibnamefont{Mazin}},
  \bibinfo{author}{\bibfnamefont{L.}~\bibnamefont{Chioncel}}, \bibnamefont{and}
  \bibinfo{author}{\bibfnamefont{A.}~\bibnamefont{Lichtenstein}},
  \bibinfo{journal}{Phys.\ Rev. B} \textbf{\bibinfo{volume}{67}},
  \bibinfo{pages}{153106} (\bibinfo{year}{2003}).

\bibitem[{\citenamefont{Anisimov et~al.}(1997)\citenamefont{Anisimov,
  Aryasetiawan, and Lichtenstein}}]{anisimov}
\bibinfo{author}{\bibfnamefont{V.}~\bibnamefont{Anisimov}},
  \bibinfo{author}{\bibfnamefont{F.}~\bibnamefont{Aryasetiawan}},
  \bibnamefont{and}
  \bibinfo{author}{\bibfnamefont{A.}~\bibnamefont{Lichtenstein}},
  \bibinfo{journal}{J.\ Phys.: Condens. Matter} \textbf{\bibinfo{volume}{9}},
  \bibinfo{pages}{767} (\bibinfo{year}{1997}).

\bibitem[{\citenamefont{Perdew et~al.}(1996)\citenamefont{Perdew, Burke, and
  Ernzerhof}}]{gga}
\bibinfo{author}{\bibfnamefont{J.}~\bibnamefont{Perdew}},
  \bibinfo{author}{\bibfnamefont{S.}~\bibnamefont{Burke}}, \bibnamefont{and}
  \bibinfo{author}{\bibfnamefont{M.}~\bibnamefont{Ernzerhof}},
  \bibinfo{journal}{Phys.\ Rev. Lett.} \textbf{\bibinfo{volume}{77}},
  \bibinfo{pages}{3865} (\bibinfo{year}{1996}).

\bibitem[{\citenamefont{Singh}(1994)}]{singh}
\bibinfo{author}{\bibfnamefont{D.}~\bibnamefont{Singh}},
  \emph{\bibinfo{title}{Planewaves, pseudopotentials and LAPW method}}
  (\bibinfo{address}{Kluwer Academic Publishers}, \bibinfo{year}{1994}).

\bibitem[{\citenamefont{Morrish}(2001)}]{morrish}
\bibinfo{author}{\bibfnamefont{A.}~\bibnamefont{Morrish}},
  \emph{\bibinfo{title}{The Physical Principles of Magnetism}}
  (\bibinfo{address}{IEEE Press, New York}, \bibinfo{year}{2001}).

\bibitem[{\citenamefont{Nagaev}(1967)}]{nagaev}
\bibinfo{author}{\bibfnamefont{E.}~\bibnamefont{Nagaev}},
  \bibinfo{journal}{JETP Letters} \textbf{\bibinfo{volume}{6}},
  \bibinfo{pages}{18} (\bibinfo{year}{1967}).

\bibitem[{\citenamefont{Gonz\'alez et~al.}(2004)\citenamefont{Gonz\'alez,
  Castro, Baldomir, Sboychakov, Rakhmanov, and Kugel}}]{ivan}
\bibinfo{author}{\bibfnamefont{I.}~\bibnamefont{Gonz\'alez}},
  \bibinfo{author}{\bibfnamefont{J.}~\bibnamefont{Castro}},
  \bibinfo{author}{\bibfnamefont{D.}~\bibnamefont{Baldomir}},
  \bibinfo{author}{\bibfnamefont{A.}~\bibnamefont{Sboychakov}},
  \bibinfo{author}{\bibfnamefont{A.}~\bibnamefont{Rakhmanov}},
  \bibnamefont{and} \bibinfo{author}{\bibfnamefont{K.}~\bibnamefont{Kugel}},
  \bibinfo{journal}{Phys. Rev. B} \textbf{\bibinfo{volume}{69}},
  \bibinfo{pages}{224409} (\bibinfo{year}{2004}).

\end{thebibliography}

\end{document}